\documentclass[%
 reprint,
 amsmath,amssymb,
 aps,
prb,
]{revtex4-2}

\usepackage{graphicx}
\usepackage{dcolumn}
\usepackage{bm}
\usepackage{hyperref}
\newcolumntype{L}{>{$}l<{$}} 
\usepackage{multirow}
\usepackage{braket}
\usepackage{comment}

\usepackage{color}

\begin{document}

\preprint{APS/123-QED}

\title{Formation of dark excitons in monolayer transition metal dichalcogenides \\ by a vortex beam: optical selection rules}

\author{Omadillo Abdurazakov}
 \email{oabdurazako@utep.edu}
\affiliation{Department of Physics, University of Texas at El Paso, El Paso, Texas 79968, USA}

\author{Chunqiang Li}%
\affiliation{Department of Physics, University of Texas at El Paso, El Paso, Texas 79968, USA}

\author{Yun-Pil Shim}%
 \email{yshim@utep.edu}
\affiliation{Department of Physics, University of Texas at El Paso, El Paso, Texas 79968, USA}

\date{\today}

\begin{abstract}
Monolayer transition metal dichalcogenides host tightly-bound excitons, which dominate their optoelectronic response even at room temperatures. Light beams are often used to study these materials with the polarization - often termed as the spin angular momentum of the light - providing the mechanism for exciting excitonic states. Light beams, however, can also carry an orbital angular momentum by creating helical structures of their phase front. In this work, we consider a Laguerre-Gaussian beam possessing an orbital angular momentum in addition to the spin angular momentum to create excitons in monolayer transition metal dichalcogenides. We derive optical selection rules that govern the allowed transitions to various exciton series using symmetry arguments. Our symmetry considerations show that we can create dark excitons using these high-order optical beams opening up new avenues for creating long-lived dark excitons with the potential of exploiting them in quantum information processing and storage. 
\end{abstract}

\maketitle

\section{\label{sec:level1} Introduction}

When a photon of a certain frequency illuminates a semiconducting crystal, in response, an exciton can be formed. During the process, an electron excited to the conduction band binds with its hole left in the valence band through the Coulomb interaction. Because ordinary light does not couple to the electron spin, it is conserved in the process. For the direct band-gap semiconductors, this kind of excitation is called a bright exciton and they are short-lived because an electron can quickly recombine with the hole emitting a photon. It is also possible for an electron to flip its spin via some nontrivial processes during the exciton formation. In contrast to the bright excitons, this kind of excitation is long-lived because the promoted electron can not  radiatively recombine without flipping its spin. Therefore, it is called a dark exciton and makes a promising candidate for a solid-state qubit due to its long recombination lifetime and coherence\cite{dark-qubit,dark-coherence} and has lately become an active field of research in its own right \cite{dark-exciton-arpes,avinash}.

The family of quasi-two-dimensional (quasi-2D) semiconducting crystals called monolayer transition metal dichalcogenides (ML TMDs) can host excitons of strikingly large binding energies compared to those of traditional semiconducting crystals such as GaAs\cite{review-tmd}. This is partly due to the weak dielectric screening and geometric confinement in quasi-2D structures. Although an indirect band-gap semiconductor in bulk, when exfoliated down to a monolayer\cite{novoselov}, a TMD crystal transitions into a direct band-gap semiconductor \cite{indirect-direct-mak,indirect-direct-splendiani}. Therefore, the optical response of these materials to light is dominated by bright exciton formation and consequent photoluminescence signal at sub band-gap frequencies. Due to the presence of heavy transition metal atoms in ML TMDs, there is a strong (moderate) spin-orbit interaction in the valence (conduction) band. Combined with the lack of spatial inversion symmetry inside the monolayer crystal, the conduction and valance bands at the $\pm K$ momentum points (or valleys) are split into two bands each that can individually host only one type of electron spin species. These bands at the two valleys are related through the time-reversal symmetry. Consequently, the electron spin and valley degrees of freedom are locked and each valley can be selectively addressed with circularly polarized light \cite{valley-original,valley-cao,valley-mak,valley-zeng}. 

Because of its valley physics and tightly bound excitons that are manifest even at room temperatures, a ML TMD crystal is an ideal platform for hosting stable and long-lived dark excitons. However, creating and controlling dark excitons in these materials is challenging as ordinary light beams do not couple to the electron spin. Nevertheless, these dark excitons can be induced to decay via radiative means by applying very strong in-plane magnetic fields\cite{dark-magnetic} or coupling to the surface plasmon polaritons\cite{dark-polariton}. Although these methods offer some pathways to access the spin-forbidden dark excitons and to control their lifetimes, they are limited to extremely high magnetic fields and extremely low temperatures. Therefore, purely optical means to access and control the dark exciton states would lead to a new practical platform for many quantum applications including quantum information technology.

The polarization of light, which is used to control individual valleys in ML TMDs, corresponds to the spin angular momentum (SAM) of light and can be transferred to matter as a mechanical torque\cite{beth}. Light has another fundamental degree of freedom called orbital angular momentum (OAM). The laser beams possessing OAM can be created with holographic phase-plates and were already realized decades ago\cite{allen}. They are often called vortex beams and have been shown to induce optical transitions that are not allowed otherwise and to strongly modify optical selection rules\cite{cuprite,bound-electron}. They have been used to transfer orbital angular momentum to macroscopic particles\cite{particle}, individual atoms\cite{picon} and electrons\cite{bound-electron}, and to create Rydberg excitons in atoms \cite{portugal}, and even momentum-forbidden dark excitons in ML TMDs\cite{Taiwan}. 

In this work, we theoretically investigate the possible optical transitions with a vortex beam in ML TMD systems. Specifically, we consider a Laguerre-Gaussian laser beam, a type of vortex beam, to create spin-forbidden dark excitons as well as bright excitons in ML TMD crystals. We derive optical selection rules for such transitions based on the symmetry of the ML TMD crystal structure, the spatial structure of the beams, and the exciton envelope functions. We show that, in the quadrupole coupling regime, B-type dark excitons with various envelope functions can be created selectively in each valley. As vortex beams with high OAM can be readily created in optics labs nowadays\cite{high-oam}, this all-optical means to create dark excitons in ML TMDs should be feasible.

The outline of the paper is as follows. First, we describe our method to identify the allowed optical transition to create excitons in these materials in Sec.\ref{sec:level1} where we consider the symmetry of the electronic bands, the spatial profile of vortex beams, and exciton envelope functions in terms of the irreducible representations of the pertinent point group the monolayer crystal belongs to. In Sec.\ref{sec:level2}, we derive and discuss optical selection rules for the bright and dark excitons. And in Sec.\ref{sec:level3}, we conclude and discuss the implication of our results to future studies. 

\section{\label{sec:level1} Methods}

\begin{figure*}[t]
\includegraphics[width=0.95\textwidth]{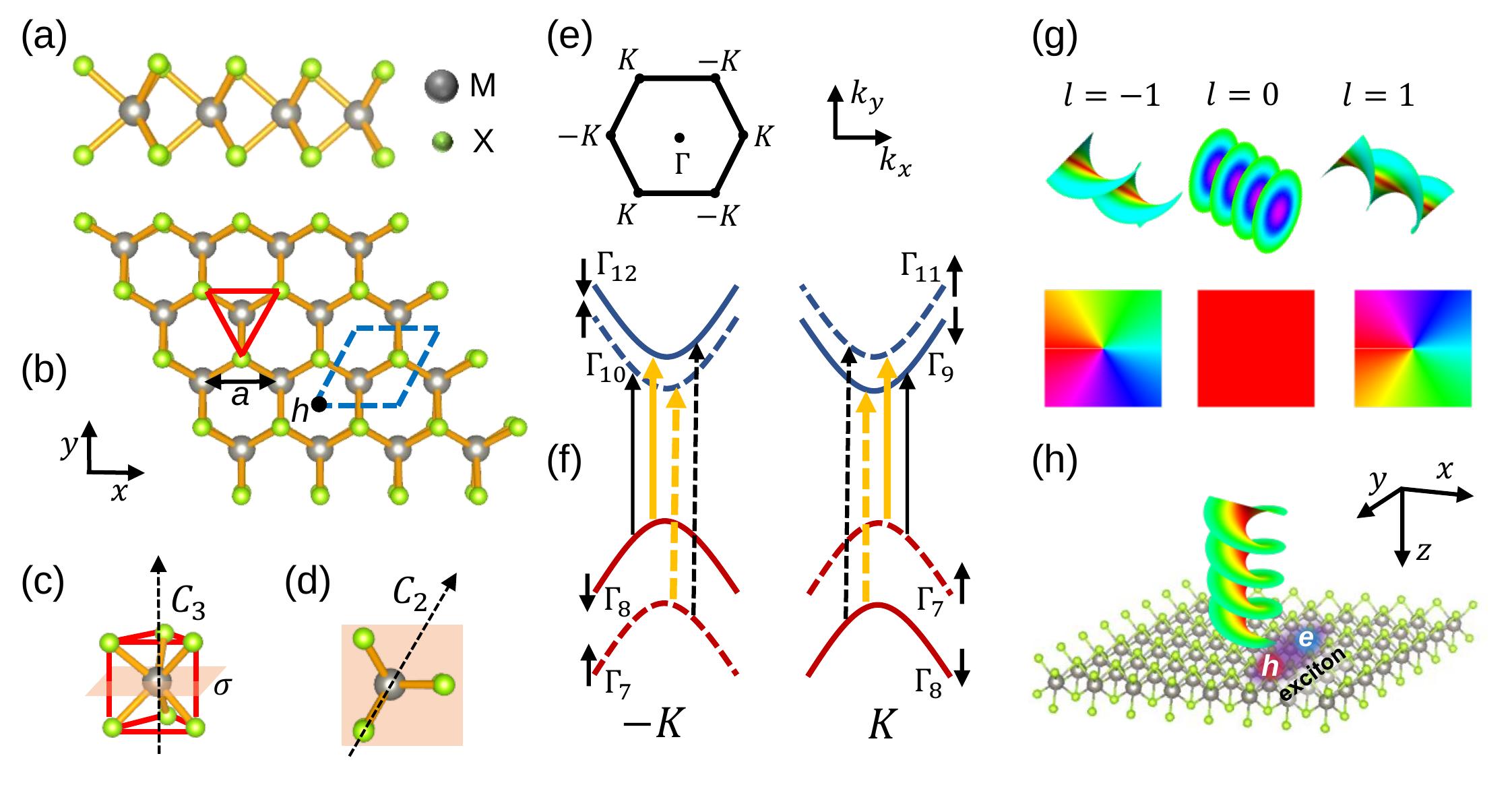}
\caption{\label{fig:one} \textbf{Monolayer transition metal dichalcogenide (ML TMD) crystal and band structure and a vortex beam.} (a) The side view and (b) top view of ML TMD crystal spanned with a unit cell (demarcated by a blue-dashed line) consisting of one transition metal and one chalcogen atom with the lattice constant of $a$.(c) The principal axis with threefold rotation symmetry, a mirror plane perpendicular to this axis, and (d) three twofold rotation axes lying on the mirror plane are shown. (e) The first Brillouin zone of ML TMD and a few high symmetry momentum points in it are shown. (f) An effective two-band model with the spin-orbit coupling splitting around $K$ and $-K$ momentum points and the corresponding irreducible representations of the double-group relevant at the $-K/K$ points. The conduction band splitting is exaggerated for clarity. The possible types of exciton transitions are symbolically shown by the arrows. Here, the solid (dashed) lines designate a A(B) type exciton. The bright (dark) excitons are shown by thick yellow (thin black) lines. Note that the arrows are shifted laterally for clarity, but the transitions predominantly occur at the valley center. (g) The helical wavefront and phases of optical vortex beams are plotted for a few values of the orbital angular momentum of light. (h) Schematic of the vortex beam created exciton in a ML TMD. The vortex beam propagates along the $z$-axis, the direction that is perpendicular to the monolayer sample.}
\end{figure*}

In this work, we are interested in understanding the exciton formation in a monolayer TMD crystal under the excitation of a vortex beam. By using symmetry arguments regarding the crystal band structure and light beam, we identify allowed and forbidden optical transitions from the ground state of the crystal to various exciton states when the light beam possesses both spin and orbital angular momenta. Such an analysis grants us the optical selection rules that govern the formation of bright and dark excitons in the $\pm K$ valleys of a monolayer TMD crystal. As the band gap energy in this class of materials predominantly lies in the optical spectrum, we only consider the direct optical transition between the spin-split valence and conduction bands at the $\pm K$ points. The symmetry of the electron wavevector around these points can be derived by constructing a tight-binding band and studying its behavior under the pertinent symmetry operations that belong to the relevant symmetry group\cite{bloch-C3}.

In the single electron band picture, an exciton with the $1s$ envelope function is formed through the binding of an electron excited into the conduction band minimum and the hole left on the valence band maximum\cite{yu-cardona}. The exciton symmetry is defined as the direct product of the irreducible representations of the exciton envelope function, the conduction band, and the valence band, namely $\Gamma_X=\Gamma_\text{env}\otimes \Gamma_c \otimes \Gamma_v^*$. For the optical transition from the crystal ground state to an exciton state to occur, $\Gamma_\text{X}$ must contain the irreducible representation of the driving field operator $\Gamma_\text{beam}=\Gamma_\text{order}\otimes \Gamma_\text{OAM} \otimes \Gamma_\text{SAM}$(see Appendix \ref{appendix:symmetry}). Here, $\Gamma_\text{order}$ is the order of the light-crystal coupling such as dipole or quadrupole. For the dipole coupling, the basis function representing its symmetry is proportional to the identity, which is represented by $\Gamma_1$ in the point group of $C_{3h}$, whereas for the quadrupole coupling the basis function is proportional to the longitudinal coordinate $z$, which is represented by $\Gamma_4$.  

To find the symmetry of the electronic bands at the $\pm K$ points, we can construct a tight-binding band and assign an irreducible representation to each band. The symmetry of the exciton envelope function is found from solutions of the 2D Hydrogen model, and the symmetry of the driving field operator is found by representing the polarization and the spatial structure of the Laguerre-Gaussian beam basis functions of the threefold rotation operator $C_3$.

In the bulk form, TMDs are indirect-gap semiconductors formed by weakly interacting layers through van-der Waals forces. A monolayer of TMD, however, is a direct-gap semiconductor with a hexagonal crystal structure. 
The side and top view of a ML TMD crystal are depicted in Figs.\ref{fig:one}(a) and (b), respectively. The crystal unit cell contains one chalcogen atom and one transition metal atom. The crystal structure is invariant under certain transformation operations. They include a threefold rotation around the principal axis perpendicular to the monolayer ($C_3$), a reflection around the mirror plane ($\sigma$) perpendicular to the principal axis, three two-fold rotations around the axes lying on the mirror plane ($C_2$), and the improper rotation around the principal axis followed by a mirror reflection ($S_3$). The set of all these operations belongs to the group of $D_{3h}$.

In the reciprocal space, the first Brillouin zone of the ML TMD crystal, which is also of a hexagonal structure, has a few high symmetry momentum points such as $\Gamma$, $K$, and $-K$ as shown in Fig.\ref{fig:one}(e). Although the electron wave function at $\Gamma$ points inherits the point group of the crystal lattice, the symmetry at $\pm K$ points is lowered to that of $C_{3h}$.  
A tight-biding wave function at these particular momentum points can be constructed from the dominant $d$ atomic orbitals of the transition metal atoms. According to the first-principle calculations \cite{tmd-dft}, the $d_{z^2}$ atomic orbitals dominate the conduction band, whereas the $d_{x^2-y^2}$ and $d_{xy}$ atomic orbitals dominate the valence band at $\pm K$ points. One can construct a tight-binding band by symmetrizing these orbital with respect to the threefold rotation operator $C_3^+$ relevant to the monolayer crystal symmetry. Here, the plus sign indicates that the three-fold rotation operator acts in the counter clockwise direction around the principal axis.  Considering the spin-orbit splitting of the valence and conduction bands at these particular momentum points, we can study the behavior of the constructed bands under the symmetry operation of the relevant double-point group using the Tab.\ref{Tab: char} and assign corresponding irreducible representation(irrep). The schematic diagrams for the bands and their irreps at $\pm K$ points are shown in Fig.\ref{fig:one}(f).

The Laguerre-Gaussian beams are high-order solutions to the paraxial wave equation in cylindrical coordinates\cite{kogelnik}. These high-order Gaussian modes exhibit more complex spatial and phase structures compared to the regular Gaussian beams. More importantly, due to the helical structure of their phase front, they carry orbital angular momentum along the propagation direction in addition to their spin angular momentum or polarization.

The spatial profile of the beam propagating along the $z$-axis in the cylindrical coordinates is given by $\bm{A}(\rho,\phi,z)=\bm{\epsilon}A_p^l(\rho,\phi,z)e^{ikz}$ where
\begin{eqnarray}
    A_p^l(\rho,\phi,z) &=& C_{pl}\frac{w_0}{w(z)}\left(\frac{\sqrt{2}\rho}{w(z)}\right)^{|l|}L_p^{|l|}
    \left( \frac{2\rho^2}{w^2(z)}\right)\nonumber\\
    &\times& e^{-\rho^2/w^2(z)}e^{i\Psi}e^{il\phi}
\end{eqnarray}
and the phase factor contains the following terms
\begin{eqnarray}
 \Psi=k\rho^2/2R(z)+(2p+|l|+1)\arctan{(z/z_R)}
\end{eqnarray}
where $C_{pl}=\sqrt{2p!/\pi(|l|+p)!}$ is the normalization constant, $R(z)=z+z_R^2/z$ is the beam front curvature, and $w(z)=w_0\sqrt{1+z^2/z_R^2}$ is the spot size which increases as a function of the distance $z$ from the beam waist $w_0$. The beam acquires a nontrivial spatial structure through the associated Laguerre function $L_p^{|l|}(\rho,z)$. The intensity profile of the beam is of a ring-like structure. Its radial index $p$, which can be zero or any positive integer, indicates the number of rings when the beam is projected on a screen. Its azimuthal index $l$, which can be any integer number, indicates the amount of orbital angular momentum the beam carries. If this index is zero, a Gaussian beam with a plane wave phase is recovered. For $p=0$ the beam intensity has a single ring structure whose radius scales as $\propto \sqrt{l}$ at the beam waist. Here, $z_R=\pi w_0^2/\lambda$ is the Raleigh range, within which the beam spot size stays nearly constant. The first term in the phase factor gives a spherical curvature to the wave, whereas the latter term is called a Gouy phase, which develops as the beam propagates and mostly varies inside the Raleigh range. And $\lambda$ is the wavelength of the light beam. The most significant phase term contains $l\phi+kz$. As depicted in Fig.\ref{fig:one}(g), the points of the constant phase make a helical surface along the $z$-axis. For $|l|>1$ there will be as many helices intertwined together. The center of the beam where the phase is undefined and the light intensity goes to zero is called a vortex. The lower panels in Fig.\ref{fig:one}(g) showcase how the phase changes as a function of the azimuthal angle $\phi$ for corresponding values of $l$. 

The light polarization vector $\bm{\epsilon}$ of the vortex beam depicted in Fig.\ref{fig:one}(h) lies parallel to the two-dimensional sample and perpendicular to the beam propagation direction. Here, the beam can be circularly polarized either in the left-handed direction ($\sigma^-$) or the right-handed direction ($\sigma^+$). It is also possible employ a beam  polarized perpendicular to the sample plane and it is termed as $\pi$-polarized beam. There are at least two ways of achieving it. First, if a Gaussian beam is directed along the two-dimensional sample plane, then its polarization is perpendicular to the sample\cite{in-plane}. Therefore, one could modify the beam-sample setup by changing the angle between them so that there is a field component that is perpendicular to the sample with a finite amount of $\pi$-polarization. Another way to create a $\pi$-polarized beam is to tightly focus the vortex beam so that it is no longer in the paraxial regime and has a longitudinal component. This non-paraxial vortex beam can posses polarization that is along the beam axis that comes from the finite longitudinal field component, albeit with much weaker amplitude. This component scales as the first order in the parameter $1/(kw_0)$ which is $\sim 10^4$ times smaller than the transverse field component for a typical vortex beam spot $w_0$ within the optical diffraction limit\cite{lax,longitudinal}. Therefore, in this work, we also consider the possibility of having a $\pi$-polarized beam for the sake of completeness in deriving the optical selection rules. 

\begin{table}[]

\begin{tabular}{|L|L|L|}
\hline
\text{Exciton env.}  & \text{Basis function} & \text{Irrep} \\ \hline
1s    &  1     &  \Gamma_1     \\ 
2p_+ &   (x+iy)    & \Gamma_2      \\ 
2p_- &   (x-iy)    & \Gamma_3      \\ 
3d_+ &  (x+iy)^2    & \Gamma_3      \\ 
3d_- &  (x-iy)^2     & \Gamma_2      \\ \hline
\end{tabular}
\caption{Basis functions and the irreps of various exction envelope functions.}
\label{Tab:env-symm}
\end{table}

Because of the quasi-2D structure of ML TMDs, the electric field lines between electrons and holes are screened well in the plane of the crystal and poorly screened outside. This renders the dielectric constant of the material dependent on the relative distance of the exciton envelope function. Therefore, to solve the exciton problem in ML TMD quantitatively one needs to resort to an elaborate numerical framework such as the Bethe-Salpeter equations\cite{BD-splitting}. However, we adopt a framework that treats the exciton problem as a 2D hydrogen model with an effective dielectric screening because we are primarily interested in the symmetry properties of the exciton envelope function rather than its exact quantitative form. It was shown that one can embed the inhomogeneous screening effects into an effective dielectric constant that is averaged over the radius of an exciton and obtain a 2D analog of the Hydrogen like solutions for exciton envelope functions with quite accurate binding energies \cite{Screened-hydrogen}. Then, the exciton envelope solutions will be of the form $\Phi_{nm}(\bm{r})=R_{n|m|}e^{im\phi}$ in real space\cite{2D-hydrogen, hydrogen-analytic}. The radial function $R$ depends on the principal quantum number $n=\{1,2,3,\ldots\}$ and the absolute value of the magnetic quantum number $m=\{0,\pm1,\ldots,\pm(n-1)\}$. It also depends on the relative coordinate of the electron and hole, the exciton radius which contains the effective dielectric constant. Because of the spherical symmetry of the radial function both in real and reciprocal space, the phase factor entirely defines the symmetry of the exciton envelope function. Because $e^{im\phi}\propto(x + (-1)^miy)^m$, one can assign an irreducible representation to each exciton state depending on the sign or the value of $m$ (see Tab. \ref{Tab:env-symm}). When $m=0$, the irrep of the envelope function is $\Gamma_1$, while for $m=\pm1$, it is $\Gamma_2$ and $\Gamma_3$, respectively. For $m>1$, it will be $n$ direct products of $\Gamma_2$ or $\Gamma_3$.

\section{\label{sec:level2} Results}

The experimental absorption measurements\cite{indirect-direct-splendiani} and first-principles calculations\cite{tmd-spectra} show that TMDs clearly exhibit A (B) excitons that are formed due to the direct optical transitions between the conduction band and spin-orbit split higher (lower) valance band. Here, we present the optical selection rules for bright A(B) and dark A(B) excitons created by vortex beams in a tabular form below. In addition to the usual dipole coupling, we also consider quadrupole coupling. The quadrupole transitions become important when the electric field has a spatial gradient. Due to their complex spatial structure, high-order laser beams such as the Laguerre-Gaussian beams, especially those with high values of OAM, have a strong special field gradient. Although the oscillator strength for the quadrupole transition is much weaker than that of the dipole transition for ordinary light beams, the vortex beams can have sizable quadrupole effects that are even comparable to those of the dipole coupling at very high values of OAM  because of their transverse field gradient around the center of the vortex\cite{quad-prl,quad-epjd}. Therefore, this second-order transition can no longer be ignored and needs a special consideration, especially for the vortex beams with large OAM. With the availability of the tools to create vortex beams with extremely high OAM\cite{high-oam}, significant quadrupole exciton transition rates are plausible. To evaluate the optical transition amplitudes and dark exciton recombination lifetimes quantitatively, one needs to resort to elaborate numerical calculations such as solving the Bethe-Salpeter equations combined with the \textit{ab-initio} band structure calculations, which will be the topics of future studies. As the vortex beams carry non-zero OAM, we also derive optical selection rules for the exciton envelope functions with higher magnetic quantum numbers complementary to the $1s$ excitons. 

\subsection{Bright Excitons}

\begin{table}[]
\begin{tabular}{|ll|LLLLLLLLLLL|}
\hline
\multicolumn{2}{|L|}{l(\text{OAM})}  & -5 & -4 & -3 & -2 & -1 & 0 & +1 & +2 & +3 & +4 & +5  \\ 
\hline
\multicolumn{1}{|L|}{\multirow{2}{*}{-K}}  & A & \oslash & \sigma^- & \sigma^+ & \oslash & \sigma^- & \sigma^+ & \oslash & \sigma^- &  \sigma^+ & \oslash & \sigma^-  \\ 
\multicolumn{1}{|L|}{}                    & B  &\oslash & \sigma^- & \sigma^+ & \oslash & \sigma^- & \sigma^+ & \oslash & \sigma^- &  \sigma^+ & \oslash & \sigma^-  \\ \hline
\multicolumn{1}{|L|}{\multirow{2}{*}{K}} & A & \sigma^+ & \oslash & \sigma^- & \sigma^+ & \oslash & \sigma^- & \sigma^+ & \oslash & \sigma^- & \sigma^+ & \oslash  \\ 
\multicolumn{1}{|L|}{}                    & B & \sigma^+ & \oslash & \sigma^- & \sigma^+ & \oslash & \sigma^- & \sigma^+ & \oslash & \sigma^- & \sigma^+ & \oslash \\ \hline
\end{tabular}
\caption{\textbf{Bright A and B excitons in the dipole coupling regime}: Optical selection rules for \textit{bright} excitons created by laser beams carrying orbital angular momentum (OAM) and spin angular momentum (SAM).}
\label{Tab:dipole-bright}
\end{table}

\begin{table}[]
\begin{tabular}{|ll|LLLLLLLLLLL|}
\hline
\multicolumn{2}{|L|}{l(\text{OAM})} & -5 & -4 & -3 & -2 & -1 & 0 & +1 & +2 & +3 & +4 & +5 \\ 
\hline
\multicolumn{1}{|L|}{\multirow{2}{*}{-K}}  & A & \pi & \oslash & \oslash & \pi & \oslash & \oslash & \pi & \oslash &  \oslash & \pi & \oslash  \\ 
\multicolumn{1}{|L|}{}                    & B & \pi & \oslash & \oslash & \pi & \oslash & \oslash & \pi & \oslash &  \oslash & \pi & \oslash \\ \hline
\multicolumn{1}{|L|}{\multirow{2}{*}{K}} & A & \oslash & \pi & \oslash & \oslash & \pi & \oslash & \oslash & \pi & \oslash & \oslash & \pi \\ 
\multicolumn{1}{|L|}{}                    & B & \oslash & \pi & \oslash & \oslash & \pi & \oslash & \oslash & \pi & \oslash & \oslash & \pi \\ \hline
\end{tabular}
\caption{\textbf{Bright A and B excitons in the quadrupole coupling regime}: Optical selection rules for \textit{bright} excitons created by laser beams carrying orbital angular momentum (OAM) and spin angular momentum (SAM).}
\label{Tab:quad-bright}
\end{table}
In the single-electron picture, the bright excitons are formed when an electron and a hole of the same spins pair via the Coulomb attraction. The optical selection rules for $1s$ bright excitons in the dipole coupling regime are presented in Tab.\ref{Tab:dipole-bright}. The OAM of the vortex beam is in the range of $|l|\leq 5$. When the beam has no OAM ($l=0$), the vortex beam reduces to the fundamental Gaussian beam with a circular polarization, and the well-known valley polarization optical selection rules are recovered\cite{valley-original}; the $\sigma^-$ polarized light can only create an exciton at the $K$ valley and the $\sigma^+$ polarized light can only create an exciton at the $-K$ valley. The two valleys and polarization directions are related through the time-reversal symmetry. In this case, the same selection rules apply to both A and B excitons. When the OAM of light $l$ is nonzero, the selection rules are modified as seen in the Tab.\ref{Tab:dipole-bright}. The allowed and disallowed transitions with $\sigma^+(\sigma^-)$ polarization alternates for a given valley and has a periodicity of three units of the OAM of light, which is inherited from the threefold rotation symmetry of the monolayer crystal. These results for bright excitons agree well with the recent study\cite{japanese} where the authors derived similar optical selection rules for the bright $1s$ excitons. However, we are mainly focused on dark excitons formation in these 2D crystals. We also show the selection rules for bright excitons in the quadrupole coupling regime in Tab.\ref{Tab:quad-bright}. We can see that we can only create excitons with particular values of $l$ using the beams that are $\pi$-polarized, i.e., directed along the light propagation axis. However, in the paraxial wave regime, the Laguere-Gauss beam is only polarized in the transverse direction. Therefore, no optical transitions are allowed in this regime for circularly polarized light. Nonetheless, one can employ tightly focused vortex beams that can have a polarization along the beam propagation axis\cite{longitudinal}.   

\subsection{Dark Excitons}

The spin-forbidden dark excitons or simply dark excitons in this work are formed when an electron and a hole with opposite spins pair via the Coulomb attraction. A light beam can not create these states because the electric field of light does not couple to the electron spin. However, the electronic bands are not purely made of one type of spin species, but mixture of both spin up and spin down species.
This spin mixing is due to the spin-orbit coupling and the interaction with remote bands\cite{in-plane}. The conduction and valence bands, therefore, contain a dominant spin component and a smaller component for the opposite spin. The transition between these bands is decided by the symmetry properties of the bands and the light.  

The optical selection rules for $1s$ dark excitons using vortex beams with $|l|\leq 5$ are presented in Tab.\ref{Tab:dipole-dark}. Here, the optical selection rules are entirely modified compared to the bright exciton cases. We can see that no circularly polarized light can excite dark excitons. Only when the beam is polarized in the direction perpendicular to the monolayer plane, one can create dark excitons. For $l=0$ only dark A-exciton can be created with the $\pi$ polarized beam that is directed parallel to the sample, which was demonstrated in photoluminescence experiments on tungsten-based TMDs\cite{in-plane}. However, the optical transition corresponding to the dark B-exciton is categorically forbidden for any polarization direction at $l=0$. For $|l|>0$ dark B excitons can be excited with $\pi$ polarized light at particular values of $l$. 

As mentioned earlier, the quadrupole transitions can be appreciable when a vortex beam is tightly focused and has a high value of OAM resulting in a strong spatial field gradient\cite{quad-prl}. Observing that the selection rules have a fixed periodicity with respect to the values of OAM, one can easily find the allowed optical transitions at very high values of the beam OAM where the quadrupole transitions may become considerable. In Tab.\ref{Tab:quad-dark}, we present the selection rules resulting from the quadrupole transitions for the A- and B-type dark excitons. Both types of dark excitons can be readily created with circularly polarized light at particular values of OAM. At $l=0$ or when the laser beam has a simple Gaussian profile, the A-type dark exciton transitions are strictly forbidden. This is the case for all values of OAM with $l\pmod{3} = 0$. With other finite OAM ($l\pmod{3} \neq 0$), A-type dark excitons can be created with circularly polarized lights, but the selection rule is the same for both $+K$ and $-K$ valleys. On the other hand, the B-type dark exciton transitions are valley-dependent. For example, it is allowed with a $\sigma^+$($\sigma^-$) polarized beam at the $K$($-K$) valley at $l=0$. Therefore, we can selectively create dark B excitons at a particular valley with circularly polarized light. This shows that the selection rules are manifestly altered in the presence of a nonzero OAM of light and starkly differ for the A and B dark excitons due to the symmetry considerations stemming from not only the band but also the spatial profile of the vortex beam, which is determined by the value of its OAM.

\begin{table}[]
\begin{tabular}{|ll|LLLLLLLLLLL|}
\hline
\multicolumn{2}{|L|}{l(\text{OAM})} & -5 & -4 & -3 & -2 & -1 & 0 & +1 & +2 & +3 & +4 & +5 \\ \hline
\multicolumn{1}{|L|}{\multirow{2}{*}{-K}} & A &\oslash & \oslash & \pi & \oslash & \oslash & \pi & \oslash & \oslash &  \pi & \oslash & \oslash \\ 
\multicolumn{1}{|L|}{}                    & B &\oslash & \pi & \oslash & \oslash & \pi & \oslash & \oslash & \pi & \oslash & \oslash & \pi \\ \hline
\multicolumn{1}{|L|}{\multirow{2}{*}{K}} & A & \oslash & \oslash & \pi & \oslash & \oslash & \pi & \oslash & \oslash & \pi & \oslash & \oslash \\ 
\multicolumn{1}{|L|}{}                    & B & \pi & \oslash & \oslash & \pi & \oslash & \oslash& \pi & \oslash & \oslash & \pi & \oslash \\ \hline
\end{tabular}
\caption{\textbf{Dark A and B excitons in the dipole coupling regime}: Optical selection rules for \textit{dark} excitons created by laser beams carrying orbital angular momentum (OAM) and spin angular momentum (SAM).}
\label{Tab:dipole-dark}
\end{table}

\begin{table}[]
\begin{tabular}{|ll|LLLLLLLLLLL|}
\hline
\multicolumn{2}{|L|}{l(\text{OAM})} & -5 & -4 & -3 & -2 & -1 & 0 & +1 & +2 & +3 & +4 & +5 \\ \hline
\multicolumn{1}{|L|}{\multirow{2}{*}{-K}} & A  &\sigma^- & \sigma^+ & \oslash & \sigma^- & \sigma^+ & \oslash & \sigma^- & \sigma^+ &  \oslash & \sigma^- & \sigma^+ \\ 
\multicolumn{1}{|L|}{}                    & B &\sigma^+ & \oslash & \sigma^- & \sigma^+ & \oslash & \sigma^- & \sigma^+ & \oslash & \sigma^- & \sigma^+ & \oslash \\ \hline
\multicolumn{1}{|L|}{\multirow{2}{*}{K}} & A  &\sigma^- & \sigma^+ & \oslash & \sigma^- & \sigma^+ & \oslash & \sigma^- & \sigma^+ &  \oslash & \sigma^- & \sigma^+ \\ 
\multicolumn{1}{|L|}{}                    & B &\oslash & \sigma^- & \sigma^+ & \oslash & \sigma^- & \sigma^+ & \oslash & \sigma^- & \sigma^+ & \oslash & \sigma^- \\ \hline
\end{tabular}
\caption{\textbf{Dark A and B excitons in the quadrupole coupling regime}: Optical selection rules for \textit{dark} excitons created by laser beams carrying orbital angular momentum (OAM) and spin angular momentum (SAM).}
\label{Tab:quad-dark}
\end{table}

\subsection{Exciton Envelope Functions}

The experimental and theoretical studies have shown that the excitonic properties of 2D TMDs differ starkly from those in the bulk. This is mainly due to non-local Coulomb screening in and out of the plane of the monolayer TMD. For example, the optical reflectance measurements in a typical ML TMD, $\text{WSe}_2$, show that the exciton series in these materials clearly deviate from the 2D Hydrogen series\cite{non-hydrogen}. The two-photon excitation spectroscopy combined with the many-body perturbation theory within the GW self-energy approximation and the Bethe-Salpeter Equations also attest to the same\cite{nature-non-hydrogen}. However, these calculations also showed that there is a one-to-one correspondence between the exciton wave function symmetry in ML TMDs and 2D Hydrogen model. By treating the exciton problem in ML TMDs as a 2D Hydrogen model for the sake of exciton wave function symmetry, we explore the higher-lying exciton states such as $2p_+$ and $2p_-$. The symmetries of the exciton states such as $3d_\pm$ and $4f_\pm$ can be reduced to the symmetries of $2p_\pm$.

\begin{figure}[t]
\includegraphics[width=0.99\columnwidth]{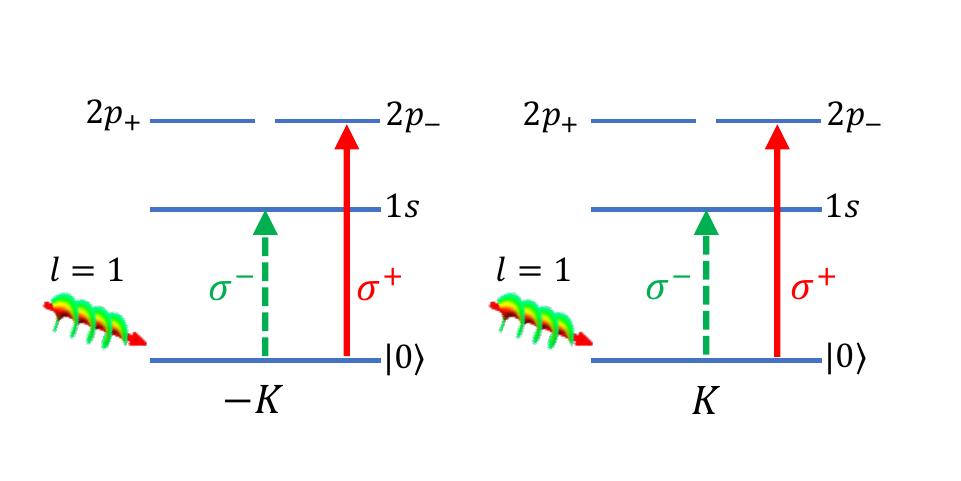}
\caption{\label{fig:two} \textbf{Dark A excitons in the quadrupole coupling regime}: Diagram describes the allowed transitions from the crystal ground state to the $1s$ and $2p_\pm$ exciton states for the dark B excitons when the vortex beam has $l=1$ orbital angular momentum. The symmetries of all higher lying exciton states can be reduces to the symmetries of these three states.}
\end{figure}

\begin{figure}[t]
\includegraphics[width=0.99\columnwidth]{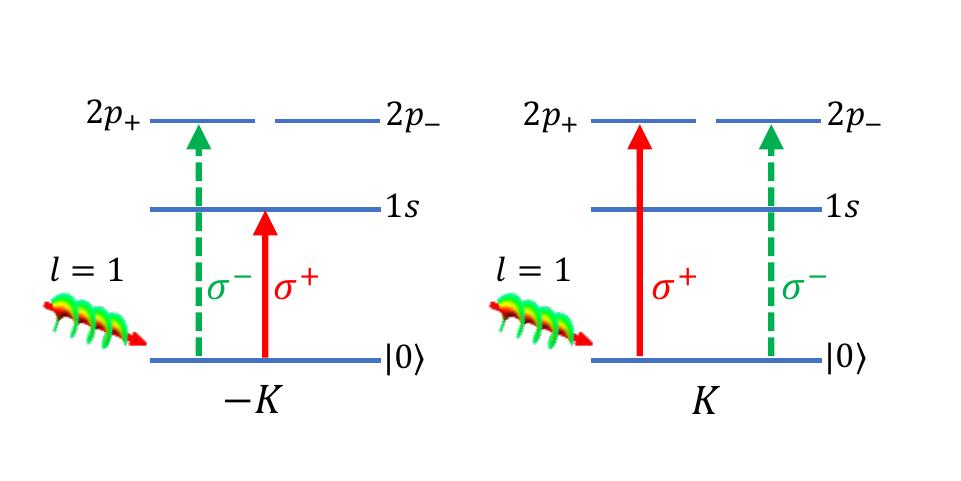}
\caption{\label{fig:three} \textbf{Dark B excitons in the quadrupole coupling regime}: Diagram describes the allowed transitions from the crystal ground state to the $1s$ and $2p_\pm$ exciton states for the dark B excitons when the vortex beam has $l=1$ orbital angular momentum. The symmetries of all higher lying exciton states can be reduces to the symmetries of these three states.}
\end{figure}

In Tab.\ref{Tab:dipole-bright-A-envelope}, we tabulated the optical selection rules for the bright A and B excitons of the $1s$ and $2p_\pm$ series in the dipole coupling regime. Both A- and B- type bright excitons follow the same selection rules. The selection rules for the $2p_\pm$ excitons differ from those of the $1s$ excitons, nevertheless, the order of the light polarization with respect to the OAM stays the same -- the pattern has the periodicity of three units of OAM. Because dark exciton states can not be realized with circularly polarized light in the dipole coupling regime(see Tab.\ref{Tab:dipole-dark}), we only display the selection rules for dark excitons in the quadrupole regime in Tab.\ref{Tab:quad-dark-A-envelope} and Tab.\ref{Tab:quad-dark-B-envelope} for the $1s$ and $2p_\pm$ series. It is possible to create dark excitons with vortex beams in the quadrupole coupling regime for various exciton states. Using an appropriate vortex beam, we can selectively excite $1s$ or $2p_\pm$ dark A (B) excitons.
In Fig.\ref{fig:two} and Fig.\ref{fig:three}, we pictorially show the allowed A and B dark exciton transitions from the crystal ground state to the $1s$, $2p_+$, and $2p_-$ exciton states for the vortex beam for a particular value of the light OAM ($l=1$). According to the diagrams, the selection rules are the same in both valleys for the dark A excitons and differ for the dark B excitons.

\begin{table}[]
\begin{tabular}{|LL|LLLLLLLLLLL|}
\hline
\multicolumn{2}{|L|}{l(\text{OAM})}                     & -5 & -4 & -3 & -2 & -1 & 0 & +1 & +2 & +3 & +4 & +5 \\ \hline
\multicolumn{1}{|L|}{\multirow{2}{*}{1s}}    & \text{-K} & \oslash & \sigma^- & \sigma^+ & \oslash & \sigma^- & \sigma^+ & \oslash & \sigma^- &  \sigma^+ & \oslash & \sigma^-   \\ 
\multicolumn{1}{|L|}{}                       & \text{K}  & \sigma^+ & \oslash & \sigma^- & \sigma^+ & \oslash & \sigma^- & \sigma^+ & \oslash & \sigma^- & \sigma^+ & \oslash  \\ \hline
\multicolumn{1}{|L|}{\multirow{2}{*}{$2p_{+}$}} & \text{-K}  & \sigma^+ & \oslash & \sigma^- & \sigma^+ & \oslash & \sigma^- & \sigma^+ & \oslash & \sigma^- & \sigma^+ & \oslash  \\ 
\multicolumn{1}{|L|}{}                       & \text{K}  & \sigma^- & \sigma^+ & \oslash & \sigma^- & \sigma^+ & \oslash & \sigma^- & \sigma^+ & \oslash & \sigma^- & \sigma^+  \\ \hline
\multicolumn{1}{|L|}{\multirow{2}{*}{$2p_{-}$}} & \text{-K}  & \sigma^- & \sigma^+ & \oslash & \sigma^- & \sigma^+ & \oslash & \sigma^- & \sigma^+ & \oslash & \sigma^- & \sigma^+ \\ 
\multicolumn{1}{|L|}{}                       & \text{K} & \oslash & \sigma^- & \sigma^+ & \oslash & \sigma^- & \sigma^+ & \oslash & \sigma^- & \sigma^+ & \oslash & \sigma^-  \\ \hline
\end{tabular}
\caption{\textbf{Bright A and B excitons in the dipole coupling regime}: Optical selection rules for the \textit{bright} A and B excitons created by laser beams carrying orbital anglar momentum (OAM) and spin angular momentum (SAM) for $1s$ and $2p_{\pm}$ exciton envelope functions.}
\label{Tab:dipole-bright-A-envelope}
\end{table}

\begin{table}[]
\begin{tabular}{|LL|LLLLLLLLLLL|}
\hline
\multicolumn{2}{|L|}{l(\text{OAM})}                      & -5 & -4 & -3 & -2 & -1 & 0 & +1 & +2 & +3 & +4 & +5 \\ \hline
\multicolumn{1}{|L|}{\multirow{2}{*}{1s}}    & \text{-K} &\sigma^- & \sigma^+ & \oslash & \sigma^- & \sigma^+ & \oslash & \sigma^- & \sigma^+ &  \oslash & \sigma^- & \sigma^+   \\ 
\multicolumn{1}{|L|}{}                       & \text{K}  &\sigma^- & \sigma^+ & \oslash & \sigma^- & \sigma^+ & \oslash & \sigma^- & \sigma^+ &  \oslash & \sigma^- & \sigma^+  \\ \hline
\multicolumn{1}{|L|}{\multirow{2}{*}{$2p_{+}$}} & \text{-K}  & \oslash & \sigma^- & \sigma^+ & \oslash & \sigma^- & \sigma^+ & \oslash & \sigma^- & \sigma^+ & \oslash & \sigma^-  \\ 
\multicolumn{1}{|L|}{}                       & \text{K}  & \oslash & \sigma^- & \sigma^+ & \oslash & \sigma^- & \sigma^+ & \oslash & \sigma^- & \sigma^+ & \oslash & \sigma^-  \\ \hline
\multicolumn{1}{|L|}{\multirow{2}{*}{$2p_{-}$}} & \text{-K} & \sigma^+ & \oslash & \sigma^- & \sigma^+ & \oslash & \sigma^- & \sigma^+ & \oslash & \sigma^- & \sigma^+ & \oslash \\ 
\multicolumn{1}{|L|}{}                       & \text{K} & \sigma^+ & \oslash & \sigma^- & \sigma^+ & \oslash & \sigma^- & \sigma^+ & \oslash & \sigma^- & \sigma^+ & \oslash \\ \hline
\end{tabular}
\caption{\textbf{Dark A excitons in the quadrupole coupling regime}: Optical selection rules for the \textit{dark} A excitons created by laser beams carrying orbital angular momentum (OAM) and spin angular momentum (SAM) for $1s$ and $2p_{\pm}$ exciton envelope functions.}
\label{Tab:quad-dark-A-envelope}
\end{table}

\begin{table}[]
\begin{tabular}{|LL|LLLLLLLLLLL|}
\hline
\multicolumn{2}{|L|}{l(\text{OAM})}                      & -5 & -4 & -3 & -2 & -1 & 0 & +1 & +2 & +3 & +4 & +5 \\ \hline
\multicolumn{1}{|L|}{\multirow{2}{*}{1s}}    & \text{-K} &\sigma^+ & \oslash & \sigma^- & \sigma^+ & \oslash & \sigma^- & \sigma^+ & \oslash & \sigma^- & \sigma^+ & \oslash  \\ 
\multicolumn{1}{|L|}{}                       & \text{K}  &\oslash & \sigma^- & \sigma^+ & \oslash & \sigma^- & \sigma^+ & \oslash & \sigma^- & \sigma^+ & \oslash & \sigma^-   \\ \hline
\multicolumn{1}{|L|}{\multirow{2}{*}{$2p_{+}$}} & \text{-K} & \sigma^- & \sigma^+ & \oslash & \sigma^- & \sigma^+ & \oslash & \sigma^- & \sigma^+ & \oslash & \sigma^- & \sigma^+ \\ 
\multicolumn{1}{|L|}{}                       & \text{K}  & \sigma^+ & \oslash & \sigma^- & \sigma^+ & \oslash & \sigma^- & \sigma^+ & \oslash & \sigma^- & \sigma^+ & \oslash   \\ \hline
\multicolumn{1}{|L|}{\multirow{2}{*}{$2p_{-}$}} & \text{-K}  & \oslash & \sigma^- & \sigma^+ & \oslash & \sigma^- & \sigma^+ & \oslash & \sigma^- & \sigma^+ & \oslash & \sigma^-  \\ 
\multicolumn{1}{|L|}{}                       & \text{K} & \sigma^- & \sigma^+ & \oslash & \sigma^- & \sigma^+ & \oslash & \sigma^- & \sigma^+ & \oslash & \sigma^- & \sigma^+  \\ \hline
\end{tabular}
\caption{\textbf{Dark B excitons in the quadrupole coupling regime}: Optical selection rules for the \textit{dark} B-excitons created by laser beams carrying orbital angular momentum (OAM) and spin angular momentum (SAM) for $1s$ and $2p_{\pm}$ exciton envelope functions.}
\label{Tab:quad-dark-B-envelope}
\end{table}

\section{\label{sec:level3} Conclusion}

Using symmetry arguments, we have derived optical selection rules for the spin-forbidden dark excitons and bright excitons in a monolayer TMD crystal excited by a vortex beam. In the dipole coupling regime, creating A and B bright excitons follows the same selection rules in individual valleys. In addition to $1s$ excitons, we can also selectively create $2p_{\pm}$ excitons. In this work, we showed the necessary conditions for creating various dark excitons in terms of the optical selection rules. Specifically, the vortex beams can be used to create A and B dark excitons in the quadrupole coupling regime and we provided the tables for optical selection rules for such transitions. This demonstrates the usefulness of the vortex beams with a nonzero orbital angular momentum in generating otherwise forbidden dark excitons, which are long-lived and have long coherence times. Optically created quantum states such as these dark exciton states can be a useful tool in many quantum device applications. For example, this could be an efficient method of transduction between semiconductor spin qubits with great tunability for information processing and photonic qubits with a long coherence time to use for communications.

\begin{acknowledgments}
C.L. acknowledges financial support from the National Science Foundation via grant NSF DMR-1827745.
\end{acknowledgments}

\appendix
\setcounter{table}{0}
\renewcommand{\thetable}{A.\Roman{table}}

\begin{table*}[]
\begin{tabular}{L|L|L|L|L|L|L|L|L|L|L|L|L|L}

C_{3h} & E & C_3^+ & C_3^- & \sigma_h & S_3^+ & S_3^- & \bar{E} & \bar{C}_3^+ & \bar{C}_3^- & \bar{\sigma}_h & \bar{S}_3^+ & \bar{S}_3^- & \text{bases}  \\ \hline
\Gamma_1 & 1 & 1 & 1 & 1 & 1 & 1 & 1 & 1 & 1 & 1 & 1 & 1 & R \\ \hline
\Gamma_2 & 1 & \omega  & \omega^* & 1 & \omega & \omega^* & 1 & \omega & \omega^* & 1 & \omega & \omega^* & x+iy \\ \hline
\Gamma_3 & 1 & \omega^*  & \omega & 1 & \omega^* & \omega & 1 & \omega^* & \omega & 1 & \omega^* & \omega & x-iy \\ \hline
\Gamma_4 & 1 & 1 & 1 & -1 & -1 & -1 & 1 & 1 & 1 & -1 & -1 & -1 & z \\ \hline
\Gamma_5 & 1 & \omega  & \omega^* & -1 & -\omega & -\omega^* & 1 & \omega & \omega^* & -1 & -\omega & -\omega^* &  \\ \hline
\Gamma_6 & 1 & \omega^* & \omega & -1 & -\omega^* & -\omega & 1 & \omega^* & \omega & -1 & -\omega^* & -\omega &  \\ \hline
\Gamma_7 & 1 & -\omega  & -\omega^* & i & -i\omega & i\omega^* & -1 & \omega & \omega^* & -i & i\omega & -i\omega^* & \chi(1/2,+1/2) \\ \hline
\Gamma_8 & 1 & -\omega^*  & -\omega & -i & i\omega^* & -i\omega & -1 & \omega^* & \omega & i & -i\omega^* & i\omega &  \chi(1/2,-1/2)\\ \hline
\Gamma_9 & 1 & -\omega & -\omega^* & -i & i\omega & -i\omega^* & -1 & \omega & \omega^* & i & -i\omega & i\omega &  \\ \hline
\Gamma_{10} & 1  & -\omega^*  & -\omega & i & -i\omega^* & i\omega & -1 & \omega^* & \omega & -i & i\omega^* & -i\omega &  \\ 
\Gamma_{11} & 1  & -1  & -1 & i & -i & i & -1 & 1 & 1 & -i & i & -i &  \\ \hline
\Gamma_{12} & 1  & -1 & -1 & -i & i & -i & -1 & 1 & 1 & i & -i & i &  \\ \hline
\end{tabular}
\caption{Character table for the $C_{3h}$ double-group\cite{koster}. Here, $\omega=\exp(i2\pi/3)$ and $\omega^*=\exp(-i2\pi/3)$.}
\label{Tab: char}
\end{table*}

\begin{table}[]
\begin{tabular}{L|LLLLLLLLLLLL}

  & \Gamma_1  & \Gamma_2  & \Gamma_3 & \Gamma_4 & \Gamma_5 & \Gamma_6 & \Gamma_7 & \Gamma_8 & \Gamma_9 & \Gamma_{10} & \Gamma_{11} & \Gamma_{12} \\\hline
\Gamma_1 & \Gamma_1 & \Gamma_2 & \Gamma_3 & \Gamma_4 & \Gamma_5 & \Gamma_6 & \Gamma_7 & \Gamma_8 & \Gamma_9 & \Gamma_{10} & \Gamma_{11} & \Gamma_{12} \\ 
\Gamma_2 & \Gamma_2  & \Gamma_3  & \Gamma_1 & \Gamma_5 & \Gamma_6 & \Gamma_4 & \Gamma_{10} & \Gamma_{12} & \Gamma_{8} & \Gamma_{11} & \Gamma_{7} & \Gamma_9 \\ 
\Gamma_3 & \Gamma_3 & \Gamma_1  & \Gamma_2 & \Gamma_6 & \Gamma_4 & \Gamma_5 & \Gamma_{11} & \Gamma_{9} & \Gamma_{12} & \Gamma_{7} & \Gamma_{10} & \Gamma_{8} \\ 
\Gamma_4 & \Gamma_4  & \Gamma_5  & \Gamma_6 & \Gamma_1 & \Gamma_2 & \Gamma_3 & \Gamma_9 & \Gamma_{10} & \Gamma_7 & \Gamma_8 & \Gamma_{12} & \Gamma_{11} \\ 
\Gamma_5 & \Gamma_5  & \Gamma_6  & \Gamma_4 & \Gamma_2 & \Gamma_3 & \Gamma_1 & \Gamma_{8} & \Gamma_{11} & \Gamma_{10} & \Gamma_{12} & \Gamma_9 & \Gamma_{7} \\ 
\Gamma_6 & \Gamma_6  & \Gamma_4  & \Gamma_5 & \Gamma_3 & \Gamma_1 & \Gamma_2 & \Gamma_{12} & \Gamma_{7} & \Gamma_{11} & \Gamma_{9} & \Gamma_8 & \Gamma_{10} \\ 
\Gamma_7 & \Gamma_7  & \Gamma_{10} & \Gamma_{11} & \Gamma_9 & \Gamma_{8} & \Gamma_{12} & \Gamma_6 & \Gamma_1 & \Gamma_3 & \Gamma_4 & \Gamma_5 & \Gamma_2 \\ 
\Gamma_8 & \Gamma_8 & \Gamma_{12} & \Gamma_{9} & \Gamma_{10} & \Gamma_{11} & \Gamma_{7} & \Gamma_1 & \Gamma_5 & \Gamma_4 & \Gamma_2 & \Gamma_3 & \Gamma_6 \\ 
\Gamma_9 & \Gamma_9 & \Gamma_{8} & \Gamma_{12} & \Gamma_7 & \Gamma_{10} & \Gamma_{11} & \Gamma_3 & \Gamma_4 & \Gamma_6 & \Gamma_1 & \Gamma_2 & \Gamma_5 \\ 
\Gamma_{10} & \Gamma_{10} & \Gamma_{11} & \Gamma_{7} & \Gamma_8 & \Gamma_{12} & \Gamma_{9} & \Gamma_4 & \Gamma_2 & \Gamma_1 & \Gamma_5 & \Gamma_6 & \Gamma_3 \\ 
\Gamma_{11} & \Gamma_{11} & \Gamma_{7} & \Gamma_{10} & \Gamma_{12} & \Gamma_9 & \Gamma_8 & \Gamma_5 & \Gamma_3 & \Gamma_2 & \Gamma_6 & \Gamma_4 & \Gamma_1 \\ 
\Gamma_{12} & \Gamma_{12} & \Gamma_9 & \Gamma_8 & \Gamma_{11} & \Gamma_{7} & \Gamma_{10} & \Gamma_2 & \Gamma_6 & \Gamma_5 & \Gamma_3 & \Gamma_1 & \Gamma_4 \\ 
\end{tabular}
\caption{Multiplication table for the $C_{3h}$ double-group\cite{koster}.}
\label{Tab: prod}
\end{table}

\section{Symmetry analysis}\label{appendix:symmetry}

\subsection{Band symmetry}

We label the electronic bands of our effective two-band (spin-split) model with their respective irreducible representations in the double-point group $C_{3h}$ relevant at the $\pm K$ momentum points. To assign a specific irrep to the valence (conduction) band we construct a tight-binding wave function by symmetry-adapted atomic orbitals with respect to the threefold rotation symmetry of the crystal. Applying the threefold rotation operator $C_3$ on this wave function gives an eigenvalue that can be used to read off their respective irreps from the character table Tab.\ref{Tab: char}.

According to the DFT band structure calculations\cite{tmd-dft}, the valence (conduction) band at the $\pm K$ momentum points in the Brillouin zone of the ML TMD crystal are predominantly made of the $d_{x^2-y^2}$ and $d_{xy}$ ($d_{z^2}$) atomic orbitals of the transition metal atoms. Using these atomic orbitals as the basis functions\cite{valley-original} that are symmetrized with respect to the threefold rotation operator $C_3^+$ relevant to the monolayer crystal symmetry, we can construct a tight-binding wave function\cite{ashcroft}. For the valence (conduction) bands denoted by $\alpha$ index at the $K$ momentum point of the reduced Brillouin zone, the wave function that satisfies the Block theorem is then given by  
\begin{eqnarray}
\psi_{\alpha}^{K}(\bm{r})=\frac{1}{\sqrt{N}}\sum_{n}e^{i\bm{K}\cdot(\bm{R}_n+\bm{\zeta})}\phi_{\alpha}(\bm{r}-\bm{R}_n-\bm{\zeta}).
\end{eqnarray}
Here, the summation is run over the lattice vectors $\bm{R}_n$, and $\bm{\zeta}=(\frac{1}{2},\frac{1}{2\sqrt{3}})a$ is the position of the transition metal atom relative to the hexagon center $h$ inside the unit cell(see Fig.\ref{fig:one}(b)). The momentum point $K=\frac{2\pi}{a}(\frac{2}{3},0)$ or the $K$ valley is take relative to  the Brillouin zone center $\Gamma=(0,0)$ (see Fig.\ref{fig:one}(e)). The valence band constructed from the atomic orbitals at this valley is $\phi_v(\bm{r})\propto \frac{1}{\sqrt{2}}(d_{x^2-y^2}+id_{xy})$ and the same for the conduction band is $\phi_c(\bm{r})\propto d_{z^2}$. 

Thus, the eigenvalue of the operator $C_3^+$ has two components: one comes from the rotation of the atomic orbital around its own center by $2\pi/3$ degrees counter-clockwise, and the other comes from the change of the Block phase after the threefold rotation. We note that the rotation operator affects the phase change differently depending on the rotation center. We took the hexagon center $h$ to be the rotation center. One could also equally take the transition metal atom to be the rotation center. However, such a choice would alter the band irreps, but the selection rules would not change. 
By applying the $C_3^+$ on the valence band at the $K$ valley $C_3^+\psi_{v}^{K}(\bm{r}) = \eta_c \psi_{v}^{K}(\bm{r})$ we obtain its eigenvalue $\eta_c = \mu_v\nu_v$, which has two components coming from the change in the Block phase $\mu_v$ and the rotation of the atomic orbital $\nu_v$. In our case, we obtain $\mu_v = e^{-i\frac{2\pi}{3}}$. The atomic orbital for the valence band can be shown to be proportional to a spherical harmonic function as $\phi_v(\bm{r})\propto Y_2^2(\bm{r})$ and $\nu_v=e^{+i\frac{2\pi}{3}}$. So the application of the $C_3^+$ operator on $Y_2^2(\bm{r})$ yields $\nu_v = e^{-i\frac{2\pi}{3}2}\equiv e^{+i\frac{2\pi}{3}}$. Consequently, the total eigenvalue $\eta_v=1$. From the character table Tab.\ref{Tab: char} we see that this eigenvalue of the $C_3^+$ corresponds to the $\Gamma_1$ irrep for the valency band at the $K$ valley without considering the spin-orbit interaction.

Similarly, we can follow the same procedure for the conduction band. The eigenvalue component coming from the Block phase change is the same in this case as well $\mu_c = e^{-i \frac{2\pi}{3}}$. The atomic orbital for the conduction band is also proportional to a spherical harmonic function $\phi_c(\bm{r})\propto Y_2^0(\bm{r})$. Therefore, the eigenvalue $\nu_c = e^{-i\frac{2\pi}{3}0}\equiv 1$. Consequently, the total eigenvalue $\eta_c = e^{-i\frac{2\pi}{3}}$. From the character table Tab.\ref{Tab: char} we see that this eigenvalue, on the other hand, corresponds to the $\Gamma_3$ irrep for the valency band at the $K$ valley without considering the spin-orbit interaction.

Taking the spin-orbit splitting of the valence band into account\cite{dresselhaus} gives the following irreps for the spin-split valence bands: $\Gamma_1 \otimes \Gamma_7 = \Gamma_7$ and $\Gamma_1 \otimes \Gamma_8 = \Gamma_8$ (see Tab.\ref{Tab: prod}). In this case, the spin-split band with a positive spin-orbit interaction energy lies higher in the band diagram [see Fig.\ref{fig:one}(f)]. Similarly, we can also label the spin-split conduction bands with their respective irreps: $\Gamma_3 \otimes \Gamma_7 = \Gamma_{11}$, $\Gamma_3 \otimes \Gamma_8 = \Gamma_9$. The order of the spin-split conduction bands can be inferred from the first-principles band structure calculations because the origin of the spin-orbit splitting in the conduction band is non-trivial and due to the competition between the minority orbitals coming from the chalcogen atom and the interaction with other nearby bands\cite{tmd-dft}. The irreps of the bands at the $-K$ valley are related to those of the bands at the $K$ valley by the time-reversal symmetry.

\subsection{Vortex beam profile}
During the analysis of the spatial profile of the beam, we make use of the cylindrical and spherical coordinates interchangeably. The vortex beam propagates along the $z$-axis, which is perpendicular to the ML TMD crystal lying on the $xy$ plane [see Fig.\ref{fig:one}(h)].
By placing a ML TMD crystal around the beam waist ($z\approx0$), we can obtain a much simplified spatial profile for the fundamental radial mode ($p=0$) of the Laguerre-Gauss beam where the complex Gouy phase vanishes. We limit our attention to the fundamental radial mode ($p=0$) as  $p$ does not affect the selection rules qualitatively. In the vicinity of the beam vortex ($\rho<<w_0$):
\begin{eqnarray}
 A^{l}(\rho,\phi,z) = \sqrt{\frac{2^{|l|+1}}{\pi |l|!}}\frac{\rho^{|l|}}{w_0^{|l|+1}}e^{il\phi}.
\end{eqnarray}
In the spherical coordinates, the vortex beam assumes the following form
\begin{eqnarray}
 A^{l}(r,\theta,\phi) &=& \sqrt{\frac{2^{|l|+1}}{\pi |l|!}}\frac{r^|l|}{w_0^{|l|+1}}(\sin\theta)^{|l|}e^{il\phi}\nonumber \\
 &=&(-1)^{l}\sqrt{\frac{2^{3|l|+2}|l|!}{(2|l|+1)!}}\frac{r^{|l|}}{w_0^{|l|}}Y_{|l|}^{ l},
\end{eqnarray}
where we have used the identity $\left(\sin \theta \right)^{|l|} e^{i l\phi}=(-1)^{l} 2^{|l|}|l|!\sqrt{4\pi/(2|l|+1)}Y_{|l|}^{l}$.  Here, $Y_{|l|}^{l}$ is a spherical harmonic function of degree (order) $|l|$ ($l$). As the spherical harmonic function $Y_{|l|}^{\pm l} \propto (x \pm iy)^{|l|}$, we can easily assign an irrep for specific values of the beam OAM $l$. For example, for $l=1$($-1$) it is $\Gamma_2$ ($\Gamma_3$) and for $l=0$ it is $\Gamma_1$. For high values of $l$ one needs to compute a direct product of $l$ number of $\Gamma_2$ or $\Gamma_3$ using the irrep multiplication table for the double-group of $C_{3h}$. 

\subsection{Transition matrix element}

Fermi's Golden rule states that a direct optical transition rate from the crystal ground state to the excited state is given by
\begin{eqnarray}
 R = \frac{2\pi}{\hbar}\sum_{X}|\braket{\Psi_{X}|\hat{H}_\text{int}|0}|^2\delta(E_{X}-E_0-\hbar\omega),
\end{eqnarray}
where $\hat{H}_\text{int}$ is the light-crystal interaction Hamiltonian and $\omega$ is the light frequency. The sum is taken over all final (exciton) states. The exciton wave function is a triple product of the electron-hole envelope function, the electron and hole Bloch wavefunctions at the conduction and valence bands, respectively\cite{elliott}
\begin{eqnarray}
 \Psi_{X}(\bm{r}_e, \bm{r}_h) = \sum_{\bm{k}} F_{nm}(\bm{k}) \psi_{c\bm{k}}(\bm{r}_e)\psi_{v\bm{k}}^{*}(\bm{r}_h)
\end{eqnarray}
where $F_{nm}(\bm{k})=R_{n|m|}(k)e^{im\phi}$ is the Fourier transform of the 2D exciton envelope function $\Phi_{nm}(\bm{r})$. The symmetry of the envelope function is entirely defined by the factor $\exp(im\phi)\propto(x+(-1)^miy)^m$. Depending on the value of $m$, it constitutes a basis for the irreps $\Gamma_2$ or $\Gamma_3$. Subsequently, the product of those function form a basis for the exciton representation
\begin{eqnarray}
\Gamma_{X} = \Gamma_\text{env}\otimes \Gamma_{c} \otimes \Gamma_{v}^{*}.
\end{eqnarray}
In other words, this triple direct product represents the symmetry of an exciton state.

In the Coulomb gauge, one can show that the interaction Hamiltonian
\begin{eqnarray}
\hat{H}_{int} &\propto& e^{ikz}\bm{r}\cdot\bm{\epsilon}A^l \\
&\approx& (-1)^{l}\sqrt{\frac{2^{3|l|+2}|l|!}{(2|l|+1)!}}\frac{r^{|l|+1}}{w_0^{|l|}}(1+ikz)Y_{1}^{s}Y_{|l|}^{ l}.
\end{eqnarray}
Here, the spherical harmonic function $Y_{|l|}^{ l}$  ($Y_{1}^{s}$) expresses the orbital (spin) angular momentum of light. The right (left) circularly-polarized light is designated with $s=+1(-1)$ values. When the light beam is polarized along the direction perpendicular to the crystal plane, $s=0$. As these functions form bases for the irreps in the double-point group $C_{3h}$ relevant at the $\pm K$ valleys, we can assign corresponding irreps depending on the values of $l$ and $s$. The symmetry properties of the spin degree of freedom are represented by $\Gamma_2$ ($\Gamma_3$) for $s=+1$ ($-1$) and $\Gamma_4$ for $s=0$ since $Y_1^{\pm 1}\propto (x\pm iy)$ and $Y_1^0\propto z$, respectively. Because the basis function $Y_{|l|}^l\propto(x +(-1)^l iy)^l$, the irreps $\Gamma_2/\Gamma_3$ or their $l$th power (i.e., their $l$ direct products) can represent the symmetry properties of this basis function. The order of the light-crystal coupling has the symmetry of $\Gamma_1$ for the dipole ($e^{ikz} \approx 1$) and $\Gamma_4$ for the quadrupole ($e^{ikz}\approx 1+ikz$) transition. Thus, the symmetry of the vortex beam is equal to
\begin{eqnarray}
\Gamma_\text{beam}=\Gamma_\text{order} \otimes \Gamma_\text{OAM} \otimes \Gamma_\text{SAM}.
\end{eqnarray}
For an optical transition from the crystal ground state to an excitonic state to occur, the transition matrix element $\braket{\Psi_{X}|\hat{H}_\text{int}|0}$ must be nonzero. For that to happen the triple direct product of the exciton state, vortex beam, and the crystal ground state must span the totally symmetric irrep of the group $C_{3h}$, which is $\Gamma_1$. The symmetry of the crystal ground state is also represented by $\Gamma_1$. Thus, the allowed exciton transitions satisfy the following direct product equation
\begin{eqnarray}
\Gamma_{X}^{*}\otimes \Gamma_\text{beam}\otimes \Gamma_1 = \Gamma_1.
\end{eqnarray}
This statement is also known as the matrix element theorem\cite{yu-cardona,hopfield}, which states that the symmetry of the final (exciton) state must contain symmetry of the driving field operator for an excitation from the ground state to occur.


\providecommand{\noopsort}[1]{}\providecommand{\singleletter}[1]{#1}%

\end{document}